# Hydrodynamics of Stellar Mergers and the Formation of Blue Stragglers

Frederic A. Rasio

*Department of Physics, MIT 6-201, Cambridge, MA 02139*

**Abstract.** The hydrodynamics of stellar collisions and mergers is discussed in the context of blue straggler formation. Emphasis is placed on the very important question of hydrodynamic mixing during the merger process. Recent results of three-dimensional hydrodynamic calculations suggest that the merger remnants produced by stellar collisions are typically *not* well-mixed. However, comparisons between the observed colors and numbers of blue stragglers in dense clusters and the predictions of theoretical calculations for their stellar evolution appear to require that the initial blue-straggler models be close to chemically homogeneous. The resolution of this apparent conflict is likely to involve the development of convection or other thermal instabilitites that can provide efficient mixing during the contraction of the merger remnant to a thermal equilibrium state on the main sequence.

## 1. Introduction

Blue stragglers are main-sequence stars that appear above the turnoff point in the color-magnitude diagram of a cluster. They are thought to be formed through a merger of two lower-mass stars, which can occur following the physical collision of two single stars or the coalescence of the two components in a close binary system (Leonard 1989; Livio 1993; Stryker 1993; Bailyn 1995).

Direct evidence for binary progenitors has been found in the form of contact (W UMa type) binaries among blue stragglers in the low-density globular clusters NGC 5466 (Mateo et al. 1990) and M71 (Yan & Mateo 1994), as well as in open clusters (Kałużny & Ruciński 1993; Milone & Latham 1994; Jahn, Kałużny & Ruciński 1995). Indications of a collisional origin come from recent detections by HST of large numbers of blue stragglers concentrated in the cores of some of the densest clusters, such as M15 (De Marchi & Paresce 1994; Guhathakurta et al. 1995) and M30 (Yanny et al. 1994), and from the apparent lack of binaries in these very dense environments (Shara et al. 1995).

Because the details of the hydrodynamic merger process are better understood theoretically in the case of stellar collisions than for binary coalescence, this paper will focus on the former. For recent discussions of binary coalescence in the context of blue straggler formation, see Rasio (1993, 1995) and Rasio & Shapiro (1995).

Collisions can happen directly between two single stars only in the cores of the densest clusters, but even in somewhat lower-density clusters they can also happen indirectly, during resonant interactions involving primordial binaries (Sigurdsson, Davies, & Bolte 1994; Sigurdsson & Phinney 1995; Davies & Benz 1995). Observational evidence for the existence of dynamically significant numbers of primordial binaries in globular clusters is now well established (Hut et al. 1992; Cote et al. 1994; various reviews in these proceedings).

## 2. The Importance of Hydrodynamic Mixing

Benz & Hills (1987, 1992) performed the first three-dimensional calculations of direct collisions between two main-sequence stars. An important conclusion of their pioneering study was that stellar collisions could lead to thorough mixing of the fluid. In particular, they pointed out that the mixing of fresh hydrogen fuel into the core of the merger remnant could reset the nuclear clock of a blue straggler, allowing it to remain visible for a full main-sequence lifetime $t_{MS} \sim 10^9$ yr after its formation.

In subsequent work it was generally assumed that the merger remnants resulting from stellar collisions were nearly homogeneous. Blue stragglers would then start their life close to the zero-age main sequence, but with an anomalously high helium abundance coming from the hydrogen burning in the parent stars. In contrast, little hydrodynamic mixing was expected to occur during the much gentler process of binary coalescence, which could take place on a stellar evolution timescale rather than on a dynamical timescale (Mateo et al. 1990; Bailyn 1992; but see Rasio 1993, 1995, and Rasio & Shapiro 1995).

On the basis of these ideas, Bailyn (1992) suggested a way of distinguishing observationally between the two possible formation processes. The helium abundance in the envelope of a blue straggler, which reflects the degree of mixing during its formation process, can affect its observed position in a color-magnitude diagram. Blue stragglers made from collisions would have a higher helium abundance in their outer layers than those made from binary mergers, and this would generally make them appear somewhat brighter and bluer.

A detailed analysis was carried out by Bailyn & Pinsonneault (1995) who performed stellar evolution calculations for blue stragglers assuming various initial chemical composition profiles. To represent the collisional case, they assumed chemically homogeneous initial profiles with enhanced helium abundances, calculating the total helium mass from stellar evolution models of the parent stars. For the dense cluster 47 Tuc they concluded that the observed luminosity function and numbers of blue stragglers were then consistent with a collisional origin.

## 3. Numerical Calculations and Stellar Models

The vast majority of recent three-dimensional calculations of stellar interactions (collisions, binary coalescence, common envelope evolution, tidal disruption, etc.) have been done using the smoothed particle hydrodynamics (SPH) method (see Monaghan 1992 for a recent review). Since SPH is a Lagrangian method, in which particles are used to represent fluid elements, it is ideally suited for the study of hydrodynamic mixing. Indeed, chemical abundances are passively advected quantities during the dynamical evolution. Therefore, the chemical composition in the final fluid configuration can be determined after the completion of a calculation simply by noting the original and final positions of all SPH particles and by assigning particle abundances according to an initial profile.

Lombardi, Rasio, & Shapiro (1995a,b) have recently re-examined the question of mixing in stellar collisions by performing a new set of numerical hydrodynamic calculations using SPH. This new work improves on the previous study of Benz & Hills (1987) by adopting more realistic stellar models, and by performing numerical calculations with increased spatial resolution. Benz & Hills (1987) used an early version of the SPH method and performed their calculations with a small number of particles ($N = 1024$). They also represented all stars by simple $n = 1.5$ polytropes. Unfortunately, $n = 1.5$ polytropes have density profiles that are not steep enough to represent main-sequence stars close to the turnoff point. Turnoff main-sequence stars have very shallow convective envelopes and are much better modeled by $n = 3$ polytropes (which have much more centrally concentrated density profiles).

The new SPH calculations of Lombardi et al. (1995a,b) are done using $N = 3 \times 10^4$ particles, and the colliding stars are modeled as composite polytropes (with $n = 3$ for the radiative interior and $n = 1.5$ in the convective envelope), which provide accurate representations of the density profiles in the entire mass range of interest for globular clusters (cf. Rappaport, Verbunt, & Joss 1983; Ruciński 1988). This is particularly important for collisions between two stars of different masses, which in general will also have different internal structures (and, for this reason, the later calculations of Benz & Hills 1992, done for two $n = 1.5$ polytropes with a mass ratio of 1/5, are of very limited applicability).

Stars close to the main-sequence turnoff point in a cluster are in fact the most relevant ones to consider for stellar collision calculations. Indeed, as the cluster evolves via two-body relaxation, the most massive stars will tend to concentrate in the dense cluster core, where the collision rate is highest (see, e.g., Spitzer 1987). In addition, collision rates can be increased dramatically by the presence of a significant fraction of primordial binaries in the cluster, and the more massive stars will preferentially tend to be exchanged into such a binary, or collide with another star, following a dynamical interaction between two binaries or between a binary and a single star (Sigurdsson & Phinney 1995).

## 4.   Summary of Recent Results

The main new results of Lombardi et al. (1995a,b) can be summarized as follows. Typical merger remnants produced by collisions are far from chemically homogeneous. For example, Fig. 1 shows the helium abundance profile inside a merger remnant formed by the collision of two $0.8\,M_\odot$ turnoff main-sequence stars. It is clear that hydrodynamic mixing during the collision has been minimal. In fact, the final chemical composition profile is very close to the initial profile of the parent stars. For two turnoff stars, this means that the core of the merger remnant is still mostly helium. Clearly, an object with a chemical composition profile like the one shown in Fig. 1 will not be able to remain on the main sequence for a very long time, since it is born with very little hydrogen left to burn at the center.

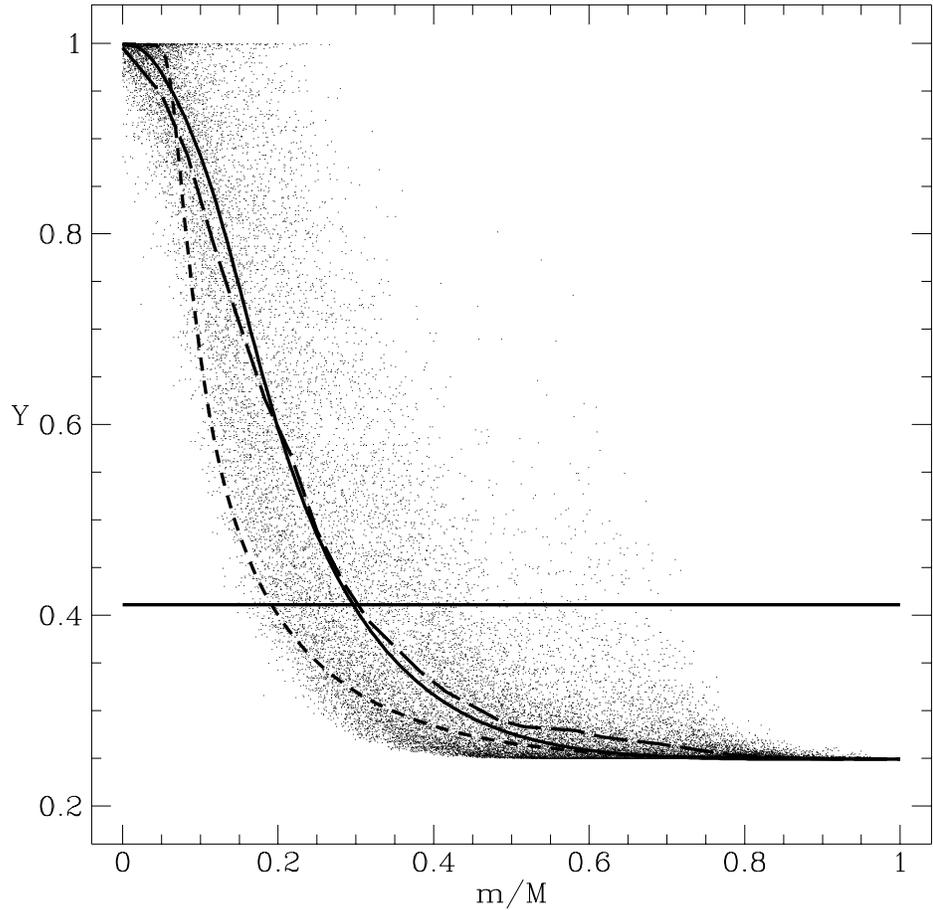

Figure 1. Helium mass fraction $Y$ as a function of interior mass fraction $m/M$ in the final configuration of a merger between two identical $0.8\,M_\odot$ turnoff main-sequence stars. The dots correspond to the individual SPH particle values at the end of the calculation, and the long-dashed curve represents their average. Also shown for comparison are the profiles corresponding to a completely mixed merger remnant (horizontal line), a terminal-age main-sequence star with the same total mass (short-dashed curve), and the initial colliding stars (solid curve). It is clear that the merger remnant is not at all well mixed, but instead has a chemical composition profile which resembles closely that of the parent stars.

In the case of collisions between two stars of very different masses, the chemical composition profiles of the merger remnants can be rather peculiar. For example, it often happens that the maximum helium abundance does not occur at the center of the remnant. This is illustrated in Fig. 2, which compares interior profiles of the final configurations following collisions between two stars with a mass ratio of 1/2 and between two identical stars, at various impact parameters.

Typical merger remnants produced by off-axis collisions are found to be rapidly and differentially rotating. The ratio of rotational kinetic energy to gravitational binding energy $T/|W|$ sometimes approaches the secular stability limit, $T/|W| \simeq 0.14$. The ratio $\Omega^2 r_e/g$ of the centrifugal to gravitational acceleration in the equatorial plane ($\Omega$ is the angular velocity, $r_e$ is the equatorial radius, and $g$ is the gravitational acceleration at the equator) can have values approaching unity in the outer layers, indicating that some configurations are rotating near break-up. The final rotating configurations are *not* barotropic (i.e., surfaces of constant density and pressure do not coincide) and, as a result, the specific angular momentum is not simply constant on cylinders centered on the rotation axis (see Fig. 3).

## 5. Discussion

At the qualitative level, the dynamical merger process in globular clusters can be understood very simply in terms of the requirement of convective stability of the final configurations. If entropy production in shocks could be neglected (which may be reasonable for the low-velocity collisions occurring in globular clusters), then one could predict the qualitative features of the final composition profile simply by observing the composition and entropy profiles of the parent stars. Convective (dynamical) stability requires that the specific entropy $s$ increase monotonically from the center to the surface ($ds/dr > 0$, cf. Fig. 2) in the final hydrostatic equilibrium configuration. Therefore, in the absence of shock-heating, fluid elements conserve their entropy and the final composition profile of a merger remnant could be predicted simply by combining mass shells in order of increasing entropy, from the center to the outside. Many of the results mentioned in the previous section follow directly.

For example, in the case of a collision between two identical turnoff stars, it is obvious why the composition profile of the merger remnant remains very similar to that of the parent stars (Fig. 1), since shock-heating is significant only in the outer layers of the stars, which contain a very small fraction of the total mass. The low-entropy, helium-rich material is concentrated in the deep interior of the parent stars, where shock-heating is negligible, and therefore it remains concentrated in the deep interior of the final configuration. For two stars of very different masses, the much lower-entropy material of the lower-mass star tends to concentrate in the core of the final configuration, leading to the unusual composition and temperature profiles seen in Fig. 2(a). In essence, the smaller-mass star simply sinks in and settles at the center of the merger, while the higher-entropy, helium-rich material has been pushed out.

Sills, Bailyn & Demarque (1995, and these proceedings) were the first to explore the consequences of blue stragglers being born unmixed. Using detailed stellar structure calculations, they compared the predicted colors ($U - B$ and $B - V$) of initially unmixed blue stragglers with observations. They concluded that some blue stragglers have observed colors that *cannot be explained* using unmixed initial models. Initially homogeneous models, however, can reproduce all the observations. In addition, unmixed models have much shorter main-sequence lifetimes than homogeneous models (because the cores of unmixed models have a very limited supply of hydrogen to burn), and therefore may be incompatible with the observed *numbers* of blue stragglers.

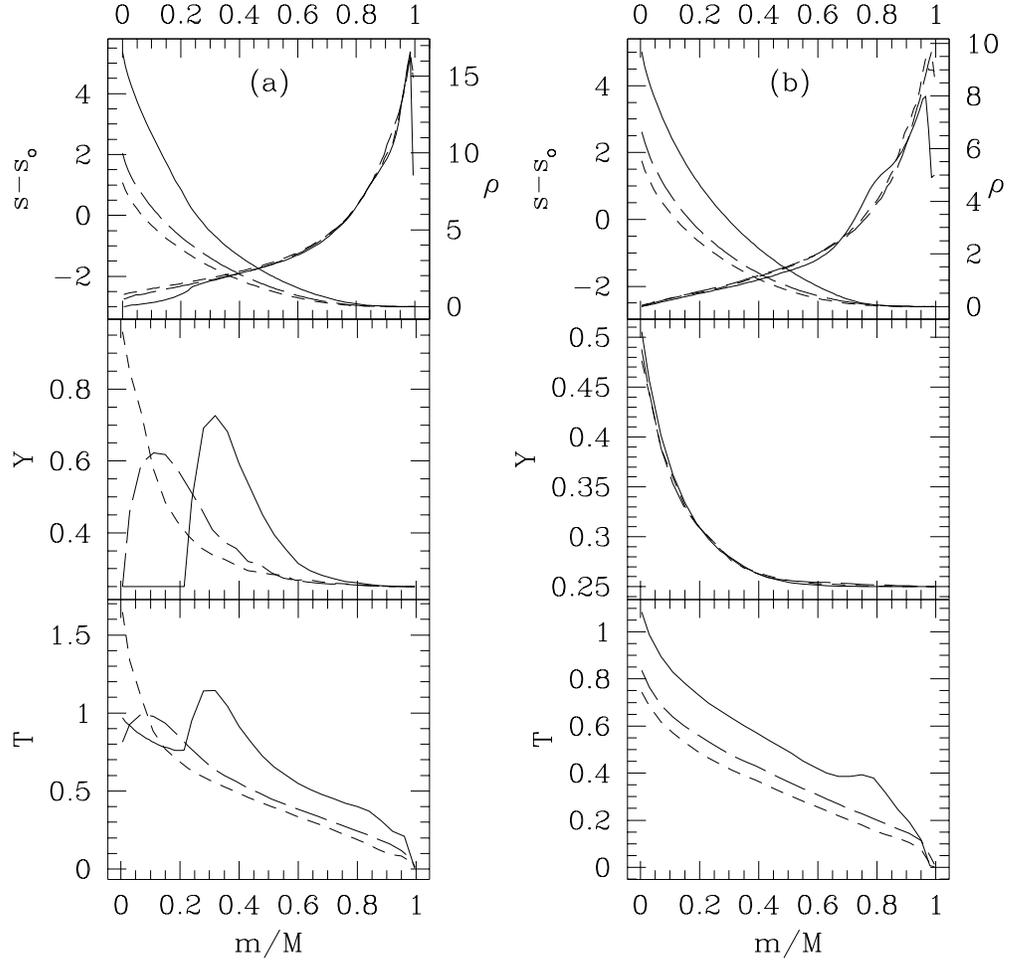

Figure 2. Final interior profiles for several merger remnants. The density $\rho$, specific entropy $s$ (up to a constant $s_o$), fractional helium abundance $Y$, and temperature $T$ are shown as a function of interior mass fraction. The masses of the two colliding stars were $M_1 = 0.8\,M_\odot$ and $M_2 = 0.4\,M_\odot$ in (a), and $M_1 = M_2 = 0.6\,M_\odot$ in (b). The solid, long-dashed and short-dashed curves correspond to initial trajectories with increasing periastron separations, $r_p/(R_1+R_2) = 0$, 0.25, and 0.5, respectively ($R_1$ and $R_2$ are the initial stellar radii). The units are defined by $G = M_{TO} = R_{TO} = 1$, where $M_{TO} \simeq 0.8\,M_\odot$ and $R_{TO} \simeq 1\,R_\odot$ (i.e., the mass and radius of a turnoff main-sequence star). The specific entropy $s$ increases monotonically from the center to the outside, as required for convective stability (except in the outermost few percent of the mass, which have not yet reached hydrostatic equilibrium). Note the peculiar composition and temperature profiles in (a).

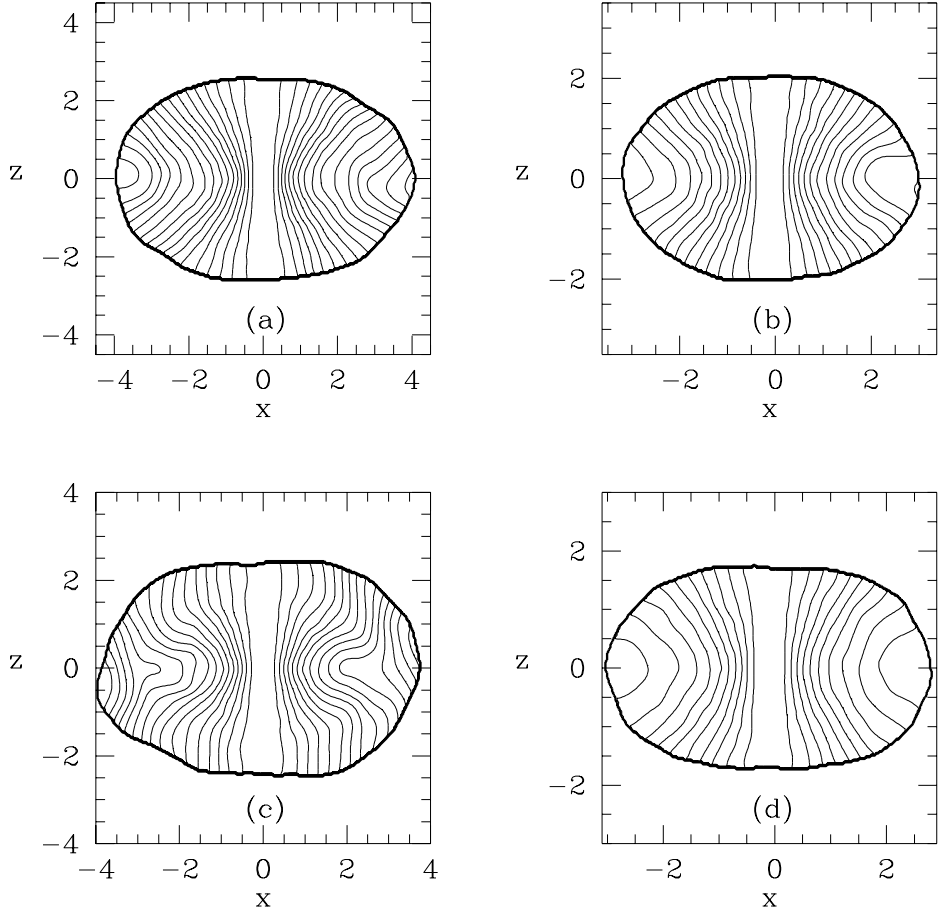

Figure 3. Contours of specific angular momentum $\Omega\varpi^2$ ($\Omega$ is the angular velocity and $\varpi$ is the cylindrical radius measured from the rotation axis) in the vertical $(x,z)$ plane (meridional section, $z$ being the rotation axis) for typical merger remnants. Units are defined as in Fig. 2. The contours have a linear spacing of $0.1\,(GM_{TO}R_{TO})^{1/2}$, with the specific angular momentum increasing away from the rotation axis (as required for dynamical stability). The thick bounding curve marks the isodensity surface which encloses 95% of the total gravitationally bound mass. In (a), the colliding stars had masses $M_1 = M_2 = 0.8\,M_\odot$ and the initial trajectory had a pericenter separation $r_p/(R_1+R_2) = 0.25$. In (b), $M_1 = 0.8\,M_\odot$, $M_2 = 0.6\,M_\odot$, and $r_p/(R_1+R_2) = 0.25$. In (c), $M_1 = M_2 = 0.6\,M_\odot$, and $r_p/(R_1+R_2) = 0.5$. In (d), $M_1 = 0.6\,M_\odot$, $M_2 = 0.4\,M_\odot$, and $r_p/(R_1+R_2) = 0.5$. In all four cases the final merger remnant has $T/|W| \simeq 0.1$.

Thus we now have some indications from the observations that *significant mixing must take place* during the blue straggler formation process, and this is in apparent conflict with the latest results of hydrodynamic calculations. In addition, spectroscopic measurements of surface rotation rates of blue stragglers indicate that they are slow rotators (Mathys 1987, 1991), also in contradiction with the predictions of dynamical merger calculations. These problems are not peculiar to stellar collisions. Binary coalescence on a dynamical timescale also produces rapidly rotating merger remnants, and since the binary coalescence process is less dissipative than direct collisions (i.e., it creates less shock-heating), we expect even less hydrodynamic mixing in this case (Rasio & Shapiro 1995).

To resolve these apparent conflicts, it may be necessary to take into account processes that have not yet been incorporated into the theoretical models. Hydrodynamic calculations are of course limited to following the evolution of mergers on a dynamical timescale ($t_{dyn} \sim$ hours for main-sequence stars) but are not capable of following processes taking place on a thermal timescale ($t_{th} \sim 10^7$ yr). The final configurations obtained at the end of hydrodynamic calculations (such as the ones illustrated in Figs. 2 and 3) are very close to hydrostatic equilibrium, but are generally *far from thermal equilibrium*. This is evident simply from the typical size of the merger remnants: the 95% mass radius at the end of the dynamical phase is typically several times the radius of a main-sequence star of the same total mass (see Fig. 3). Thus the object will need to contract (on its Kelvin time) before it can become a main-sequence star. In addition, the interior profiles of the merger remnants show clear evidence of departures from thermal equilibrium. A temperature gradient inversion (as seen in Fig. 2a) is a particularly clear sign. But more importantly, any rapidly rotating configuration in which $\partial\Omega/\partial z \neq 0$ (Fig. 3) must in general be out of thermal equilibrium (see, e.g., Tassoul 1978).

As the merger remnant contracts to the main sequence and evolves towards thermal equilibrium, many processes can lead to additional mixing of the fluid. These include convection (which is well known to occur during the evolution of ordinary pre-main-sequence stars), and, for rapidly rotating configurations, meridional circulation. These processes can lead not only to mixing, but also to loss of angular momentum and rapid spin-down through magnetic breaking (Leonard & Livio 1995).

Even in regions where the convective (dynamical) stability criterion $ds/dr > 0$ is satisfied, *local thermal instabilities* (i.e., secular instabilities) can still occur. The small vertical oscillations (at the local Brunt-Väisälä frequency $\Omega_{BV} \propto [ds/dr]^{1/2}$) of a fluid element in such a region have amplitudes that grow unstably on a timescale comparable to the local radiative damping time (see, e.g., Kippenhahn & Weigert 1990). In a thermally unstable region, mixing will occur on this timescale. For example, in a region where there is a positive molecular weight gradient ($d\mu/dr > 0$) stabilized by a positive temperature gradient ($dT/dr > 0$), as seen in Fig. 2 (solid lines in Fig. 2a), fingers of helium-rich material will tend to develop and penetrate the hydrogen-rich material below. Detailed calculations of the thermal relaxation phase ("pre-main-sequence blue straggler" evolution) incorporating a treatment of all relevant mixing processes will be necessary in order to develop a complete theoretical understanding of blue straggler formation.

**Acknowledgments.** The author thanks Charles Bailyn, Peter Leonard, Mario Mateo, Alison Sills and Frank Shu for useful comments. This work was supported in part by NASA Grant HF-1037.01-92A when the author was a Hubble Fellow at the Institute for Advanced Study in Princeton. Computations were performed at the Cornell Theory Center, which receives major funding from the NSF and IBM Corporation, with additional support from the New York State Science and Technology Foundation and members of the Corporate Research Institute.

**References**

Bailyn, C. D. 1992, ApJ, 392, 519

Bailyn, C. D. 1995, ARAA, 33, 133

Bailyn, C. D., & Pinsonneault, M. H. 1995, ApJ, 439, 705

Benz, W., & Hills, J. G. 1987, ApJ, 323, 614

Benz, W., & Hills, J. G. 1992, ApJ, 389, 546

Cote, P., Welch, D. L., Fischer, P., Da Costa, G. S., Tamblyn, P., Seitzer, P., & Irwin, M. J. 1994, ApJS, 90, 83

Davies, M. B., & Benz, W. 1995, MNRAS, in press

De Marchi, G., & Paresce, F. 1994, ApJ, 422, 597

Guhathakurta, P., Yanny, B., Schneider, D. P., & Bahcall, J. N. 1995, AJ, in press

Hut, P., McMillan, S., Goodman, J., Mateo, M., Phinney, E. S., Pryor, C., Richer, H. B., Verbunt, F., & Weinberg, M. 1992, PASP, 104, 981

Jahn, K., Kalużny, J., & Ruciński, S. M. 1995, A&A, 295, 101

Kalużny, J., & Ruciński, S. M. 1993, in ASP Conf. Ser. Vol. 53, Blue Stragglers, ed. R. A. Saffer (San Francisco: ASP), 164

Kippenhahn, R., & Weigert, A. 1990, Stellar Structure and Evolution (Springer-Verlag)

Leonard, P. J. T. 1989, AJ, 98, 217

Leonard, P. J. T., & Livio, M. 1995, ApJL, 447, 121

Livio, M. 1993, in ASP Conf. Ser. Vol. 53, Blue Stragglers, ed. R. A. Saffer (San Francisco: ASP), 3

Lombardi, J. C., Jr., Rasio, F. A., & Shapiro, S. L. 1995a, ApJ, 445, L117

Lombardi, J. C., Jr., Rasio, F. A., & Shapiro, S. L. 1995b, ApJ, submitted

Mateo, M., Harris, H. C., Nemec, J., & Olszewski, E. W. 1990, AJ, 100, 469

Mathys, G. 1987, A&AS, 71, 201

Mathys, G. 1991, A&A, 245, 467

Milone, A. A. E., & Latham, D. W. 1994, AJ, 108, 1828

Monaghan, J. J. 1992, ARAA, 30, 543

Rappaport, S., Verbunt, F., & Joss, P. C. 1983, ApJ, 275, 713

Rasio, F. A. 1993, in ASP Conf. Ser. Vol. 53, Blue Stragglers, ed. R. A. Saffer (San Francisco: ASP), 196

Rasio, F. A. 1995, ApJ, 444, L41

Rasio, F. A., & Shapiro, S. L. 1995, ApJ, 438, 887

Руciński, S. M. 1988, AJ, 95, 1895

Shara, M. M., Drissen, L., Bergeron, L. E., & Paresce, F. 1995, ApJ, 441, 617

Sigurdsson, S., Davies, M. B., & Bolte, M. 1994, ApJ, 431, L115

Sigurdsson, S., & Phinney, E. S. 1995, ApJS, 99, 609

Sills, A. P., Bailyn, C. D., Demarque, P. 1995, ApJL, in press

Spitzer, L. 1987, Dynamical Evolution of Globular Clusters (Princeton: Princeton Univ. Press)

Stryker, L. L. 1993, PASP, 105, 1081

Tassoul, J. 1978, Theory of Rotating Stars (Princeton: Princeton Univ. Press)

Yan, L., & Mateo, M. 1994, AJ, 108, 1810

Yanny, B., Guhathakurta, P., Schneider, D. P., & Bahcall, J. N. 1994, ApJ, 435, L59